# Stern-Volmer Modeling of Steady-State Forster Energy Transfer Between Dilute, Freely Diffusing Membrane-Bound Fluorophores

Jeffrey T. Buboltz, Charles Bwalya, Santiago Reyes and Dobromir Kamburov
*Department of Physics and Astronomy, Colgate University, Hamilton, New York, 13346*

Two different metrics are used to assess Forster resonance energy transfer (FRET) between fluorophores in the steady state: (i) acceptor-quenching of donor fluorescence, $E$ (a.k.a. transfer efficiency); and (ii) donor-excited acceptor fluorescence, $F_A^{Dex}$. While $E$ is still more widely used, $F_A^{Dex}$ has been gaining in popularity for practical reasons among experimentalists who study biomembranes. Here, for the special case of membrane-bound fluorophores, we present a substantial body of experimental evidence that justifies the use of simple Stern-Volmer expressions when modeling either FRET metric under dilute-probe conditions. We have also discovered a dilute-regime correspondence between our Stern-Volmer expression for $E$ and Wolber and Hudson's series approximation for steady-state Forster quenching in 2D. This novel correspondence allows us to interpret each of our 2D quenching constants in terms of both (i) an effective Forster distance, and (ii) two maximum acceptor-concentration limits, each of which defines its own useful experimental regime. Taken together, our results suggest a three-step strategy toward designing more effective steady-state FRET experiments for the study of biomembranes.

## INTRODUCTION

The phenomenon of Förster resonance energy transfer (FRET) between two fluorescent chromophores is widely employed for a variety of purposes.[1,2] In this form, FRET may perhaps be best known as a "spectroscopic ruler," serving in applications that exploit the famous inverse sixth-order distance dependence[3] of the so-called transfer efficiency, $E$. This particular FRET metric is the fractional decrease in donor fluorescence due to acceptor quenching,[a]

$$E = 1 - \frac{F_D'}{F_D} = 1 - q_r \qquad [1]$$

so the determination of $E$ requires two separate measurements: donor fluorescence in the presence and absence of acceptor.

During the last decade, FRET has seen increasing application in studies of biological membranes, both in model systems[4] and living cells.[2] In these studies, membranes are labeled with two populations of membrane-associated fluorophores (i.e., donor and acceptor probes), and the observed FRET signal is interpreted in terms of either membrane phase behavior or specific interactions between membrane components.

Although most of these biomembrane FRET studies have been based on measurements of $E$, others have chosen to use an alternative metric: donor-excited acceptor fluorescence, $F_A^{Dex}$. This FRET metric offers three distinct advantages that appeal to the experimentalist: (i) Only a single measurement is needed (i.e., it is not necessary to prepare two parallel samples labeled with either donor alone or donor + acceptor); (ii) $F_A^{Dex}$ measurements tend to be more robust,[5] due to the sample-to-sample variation effects intrinsic to measurements of $E$; and (iii) whereas measurements of $E$ are sensitive to variations in acceptor concentration only: $E(\chi_A)$, $F_A^{Dex}$ measurements are sensitive to variations in both probe concentrations: $F_A^{Dex}(\chi_D, \chi_A)$.

In order to provide for the interpretation of experimental results, freely-diffusing probe studies must resort to a theoretical framework in order to relate variations in the FRET metric to variations in probe distributions. Whereas a common $E(\chi_D)$ framework has long been in use for membrane studies,[6,7,8] no such common model has yet emerged for $F_A^{Dex}(\chi_D, \chi_A)$. Studies based on $F_A^{Dex}$ have therefore employed phenomenological models or else resorted to computer simulation.[5]

In the experiments presented here, we have explored the utility of simple Stern-Volmer (S-V) probe-dependence expressions—for both $F_A^{Dex}(\chi_D, \chi_A)$ and $E(\chi_D)$—using six different combinations of FRET probes and membrane environments. Of course, analyses based on the S-V model are normally applied only to experiments involving collisional quenching. But dilute acceptor concentration is one condition under which Forster kinetics are known to approach the Stern-Volmer limit,[b] so under these circumstances it is also reasonable to employ an S-V model to describe FRET.

Our original goal was simply to evaluate the useful limits of an S-V expression for $F_A^{Dex}(\chi_D, \chi_A)$ so that we could use it in our FRET-based studies of membrane phase behavior.[10] However, over the course of our research we have discovered that S-V expressions can safely be used to describe both FRET metrics within acceptor-concentration ranges that are conveniently defined by an easily measured parameter: the Stern-Volmer quenching constant. Moreover, we have seen that S-V predictions can even work well up to remarkably

[a] $F_D$ is the intensity of donor fluorescence in the absence of acceptor and $F_D'$ is donor fluorescence in the presence of acceptor, so that $q_r$ is known as the *relative* donor fluorescence.

[b] The other condition being "statistical mixing" of excited-state donors and acceptors due to rapid diffusion or excitation migration. [ref 9]

high acceptor concentrations—in excess of 1.0 mole%, in some cases.

Our observations led us to compare our simple Stern-Volmer treatment with the well-known Wolber-Hudson (W-H) analysis of 2D-FRET.[7] Since the S-V and W-H models generate alternative expressions for $E(\chi_D)$, we have been able to compare their predictive power by fitting them both to experimentally determined FRET-titration curves. Our probe-titration results show that both models describe well FRET-titration data at moderate transfer efficiencies ($E < 0.6$), but at higher acceptor concentrations both models fail—each in its own characteristic way. Given that S-V expressions are considerably more convenient to work with than the W-H series approximation, we have concluded that Stern-Volmer expressions—informed by a novel interpretation of the 2D quenching constant that we have drawn from Wolber & Hudson's result in the dilute-probe limit—can be a suitable framework for interpreting FRET experiments between freely diffusing membrane-bound probes.

## STERN-VOLMER MODEL FOR FRET

Since our primary interest is in regimes of low-probe concentration, we will not concern ourselves with higher-order quenching effects and will assume a simplest possible kinetic model for the energy transfer process. Under conditions of constant donor excitation intensity at frequency $\nu_0$, we have, therefore, the following four-step model,

$$(h\nu_0+)D^o \xrightarrow{k_{De}} D^* \xrightarrow{k_{Dd}} D^o(+h\nu_1)$$
$$D^*+A^o \xrightarrow{k_{Ae}} D^o + A^*$$
$$A^* \xrightarrow{k_{Ad}} A^o(+h\nu_2)$$

in which the superscripts * and $^0$ indicate the excited and ground states; $k_{De}$ and $k_{Dd}$ are rate coefficients for donor excitation and de-excitation (all modes, including fluorescence at frequency $\nu_1$); $k_{Ae}$ (the bimolecular rate coefficient) characterizes the rate of spontaneous energy transfer; and $k_{Ad}$ describes the rate of acceptor fluorescence at frequency $\nu_2$ (together with all other first-order modes of acceptor de-excitation).

The first three steps of the model lead to a steady-state expression relating $\chi_{D*}$ to the overall probe concentrations, $\chi_D$ and $\chi_A$,[c]

$$\chi_{D*}(\chi_D,\chi_A) \approx \frac{\left(k'_{De}/k_{Dd}\right)\chi_D}{1+\left(k_{Ae}/k_{Dd}\right)\chi_A} \quad [2]$$

and together, Eqs. 1 & 2 lead to a simple expression for transfer efficiency, where $C_1 \equiv k_{Ae}/k_{Dd}$ is the so-called quenching constant:

$$E(\chi_A) = 1 - \frac{1}{1 + C_1 \cdot \chi_A} \quad [3].$$

Eq. 3 will be recognized as alternative form of the Stern-Volmer Equation,

$$q_r^{-1} = 1 + C_1 \cdot \chi_A \quad [4]$$

which is routinely used for the analysis of collisional quenching data.

Adding the fourth step generates an expression relating $\chi_{A*}$ to $\chi_D$ and $\chi_A$ in the steady state,[d]

$$\chi_{A*}(\chi_D,\chi_A) \approx \frac{\left(\frac{k'_{De}k_{Ae}k_{Ad}}{k_{Dd}}\right)\chi_D\chi_A}{1+\left(k_{Ae}/k_{Dd}\right)\chi_A}$$

which provides the S-V expression for the other FRET metric, donor-excited acceptor fluorescence:

$$F_A^{Dex}(\chi_D,\chi_A) = \frac{C_0 \cdot \chi_D \cdot \chi_A}{1 + C_1 \cdot \chi_A} \quad [5].$$

Here, $C_0$ is an instrument-specific parameter that relates an excited-state acceptor concentration to an observed $F_A^{Dex}$ for some particular experimental arrangement:

$$C_0 \equiv \left(\frac{F_A^{Dex}}{\chi_{A*}}\right)\left(\frac{k'_{De}k_{Ae}k_{Ad}}{k_{Dd}}\right).$$

Eq. 5 also implies that the interpretation of $F_A^{Dex}$ studies may be simplified further in experiments conducted at suitably low acceptor concentrations. In other words, we may define a Linear Stern-Volmer (LSV) regime in which donor-excited acceptor fluorescence is simply proportional to the product of donor and acceptor probe concentrations, and the boundary of this regime can be defined conveniently in terms of the S-V quenching constant:

$$F_A^{Dex}(\chi_D,\chi_A) \approx C_0 \cdot \chi_D \cdot \chi_A \text{ when } \chi_A << 1/C_1 \quad [6].$$

## MATERIALS AND METHODS

DPPC, DMPC and cholesterol were purchased from Avanti Polar Lipids and purity was confirmed by thin layer chromatography on washed, activated silica gel plates.[11] Both of the dialkylcarbocyanine dyes, 3,3'-dioctadecyl-oxacarbocyanine (DiO) and 1,1'-dioctadecyl-3,3,3',3'-tetramethylindocarbocyanine (DiI), were from Invitrogen Corp. (Carlsbad, CA). The fluorescent cholesterol analog, dehydroergosterol (DHE), was from Sigma-Aldrich Corp., and PIPES buffer and disodium EDTA were from Fluka Chemie AG. Aqueous buffer (2.5mM PIPES pH 7.0, 250mM KCl, 1mM EDTA) was prepared from 18 MΩ water (Barnstead E-Pure) and filtered through a 0.2 µm filter before use.

Specified sample compositions ($4.5\times10^{-7}$ moles total lipid per sample) were prepared in 13 x 100 mm screw cap tubes by combining appropriate volumes of chloroform-based lipid and probe stock solutions using gastight Hamilton volumetric syringes. 1.2 ml of aqueous buffer was then added to each tube, and the chloroform was removed by a modified version of the rapid solvent exchange procedure.[12] Samples were sealed under argon, placed in a temperature controlled water bath at 45.0°C, and then slowly cooled (~ 4°C/hour) to the appropriate temperature (i.e., 13.0°C, 23.0°C or 33.0°C), where they were held for two days before measurement. For measurements of $E$, donor mole fractions were fixed at either $\chi_{DHE} \approx 3\times10^{-3}$ or $\chi_{DiO} \approx 3\times10^{-4}$.

[c] For moderate illumination intensity (i.e., $\chi_{D*}/\chi_D << 1$), making the substitutions $k'_{De} \equiv k_{De}[h\nu_0]$, $\chi_D = \chi_{D^0} + \chi_{D*}$, and $\chi_A = \chi_{A^0} + \chi_{A*}$.

[d] Assuming both $\chi_{D*}/\chi_D << 1$ and $\chi_{A*}/\chi_A << 1$ and making the same substitutions as before.

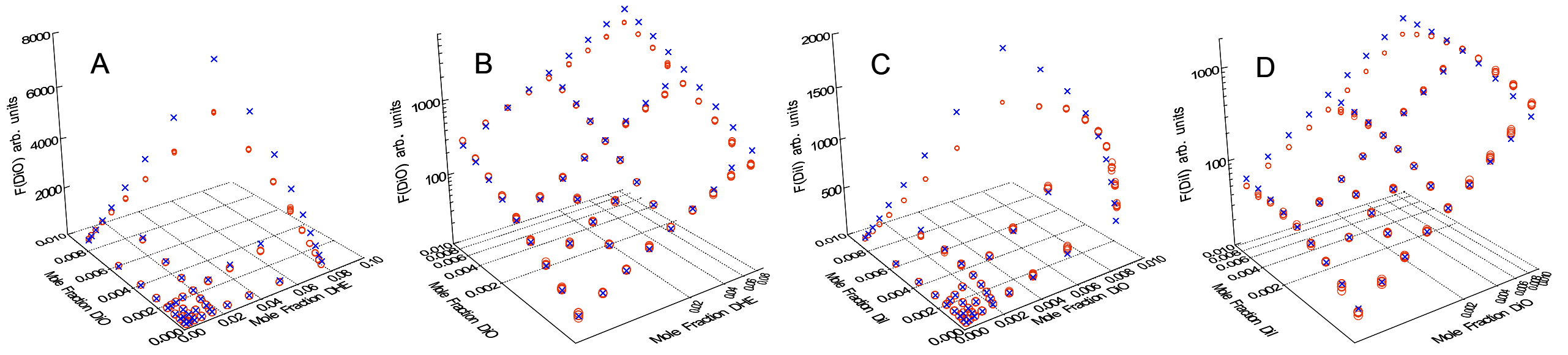

FIG 1. **Donor-excited acceptor fluorescence, $F_A^{Dex}$, exhibits Stern-Volmer behavior in the dilute-probe regime.** $F_A^{Dex}(\chi_D, \chi_A)$ titrations are shown for an $L_\beta$ gel-phase membrane environment: DMPC membranes at 13°C. In panels A and B, DHE-excited DiO fluorescence is plotted vs. probe concentration on both linear (A) and log scales (B) for comparison of high and low-probe regime behavior. Whereas panel A reveals the divergence from S-V behavior at higher probe concentrations, panel B shows clearly that $F_{DiO}^{DHEex}$ data (○) conform closely to Eq. 5 (x) in the dilute-probe regime ( $\chi_{DiO} < 2\times10^{-3}$, $\chi_{DHE} < 2\times10^{-2}$ ). For this probe combination in this membrane environment, best-fit parameter values obtained in the dilute-probe regime were $C_0 = 1.59\times10^7$ and $C_1 = 71.4$. In panels C and D, DiO-excited DiI fluorescence is plotted vs. probe concentration on linear and log scales. Again, panel C shows the divergence at higher probe concentrations, while panel D demonstrates that $F_{DiI}^{DiOex}$ data conform closely to Eq. 5 in the dilute-probe regime ( $\chi_{DiI} < 2\times10^{-3}$, $\chi_{DiO} < 2\times10^{-3}$ ) and. Dilute-probe regime best-fit parameter values: $C_0 = 2.22\times10^8$, $C_1 = 990$.

Fluorescence measurements were carried out on a Hitachi F4500 fluorescence spectrophotometer in photometry mode (10.0 sec integration; 5.0/10.0 mm slits) using a temperature-controlled cuvette holder (Quantum Northwest, Inc), and all samples were kept under argon throughout the procedure. For measurements of $F_A^{Dex}$, excitation/emission channels were set to either 325/505nm (DHE→DiO) or 430/570nm (DiO→DiI). For measurements of $E$, excitation/emission channels were set to either 325/394nm (DHE→DiO) or 430/505nm (DiO→DiI). Accurate spectral deconvolution is essential for donor-excited acceptor fluorescence measurements, so meticulous background and bleed-through corrections[5] were provided for. In brief, the F4500 was set up to record four channel combinations for each sample: a scattering signal (430/430nm) and three separate fluorescence signals ( $F_D^{Dex}, F_A^{Dex}, F_A^{Aex}$ ). Calibration standards (i.e., probe-free and single-probe samples) were included in every set of measurements, and periodic closed-shutter integrations were collected for dark current correction. After the raw fluorescence data had been corrected for each possible form of background signal (*i.e.*, dark current, scattering and spurious fluorescence), spectral deconvolution was performed, with the calibration standards serving as quality control samples. Given the wide range of probe concentrations employed in these experiments, inner-filter effect corrections were applied following deconvolution.

Least-squares fitting to probe-titration data (Figs. 1, 2 and 4) was done using a commercially available software package (Systat 11, Systat Software, Inc).

## MAPPING THE LIMITS OF STERN-VOLMER BEHAVIOR

The donor-excited acceptor fluorescence ( $F_A^{Dex}$ ) titration data shown in Fig. 1 illustrate the close agreement between the form of Eq. 5 and experimentally determined $F_A^{Dex}$ data in low-probe regimes. In order to evaluate the generality of this agreement, data were collected for two different probe pairs (DHE→DiO and DiO→DiI) in three dissimilar membrane environments: the disordered-fluid $L_\alpha$ phase; the fluid-ordered $L_o$ phase; and the gel-like $L_\beta$ phase. Our goal in these experiments was to define the limits of S-V behavior, so titrations were carried up to high probe concentrations in order to trace the eventual divergence between $F_A^{Dex}$ observations and the predictions of Eq. 5. This divergence occurs earliest and most dramatically in the gel-like $L_\beta$ phase, so these titration results are shown in Fig. 1. Even in this high-viscosity[e] membrane environment, DiO→DiI titration data still show S-V behavior throughout a dilute regime defined by $\chi_{DiI} < 2\times10^{-3}$ and $\chi_{DiO} < 2\times10^{-3}$, limits which are several-fold higher than the DiO and DiI concentrations that we routinely use in our FRET-based studies of membrane phase behavior.[10,13] The DHE→DiO $F_A^{Dex}$ data obey S-V predictions over an even wider range of probe concentrations: $\chi_{DiO} < 2\times10^{-3}$ and $\chi_{DHE} < 2\times10^{-2}$.

Since Stern-Volmer modeling should apply equally well to acceptor-quenching of donor fluorescence, $E(\chi_A)$, we also carried out standard fluorescence quenching experiments in order to determine the limits of S-V behavior for this FRET metric. Fig. 2 shows plots of relative donor fluorescence vs. acceptor concentration for the same two probe pairs in all three membrane environments. DiO→DiI data (Fig. 2B) were found to be linear up to $\chi_{DiI} \approx 1.2\times10^{-3}$, in accordance with our expectations based on $F_A^{Dex}(\chi_D, \chi_A)$ titrations. However, we found that DHE→DiO transfer efficiency obeys Stern-Volmer predictions up to acceptor concentrations in excess of 1.0 mole% acceptor (Fig. 2A), a surprisingly wide range of S-V behavior.

[e] Whereas low-viscosity environments tend to favor statistical mixing, high-viscosity environments do not [ref 9].

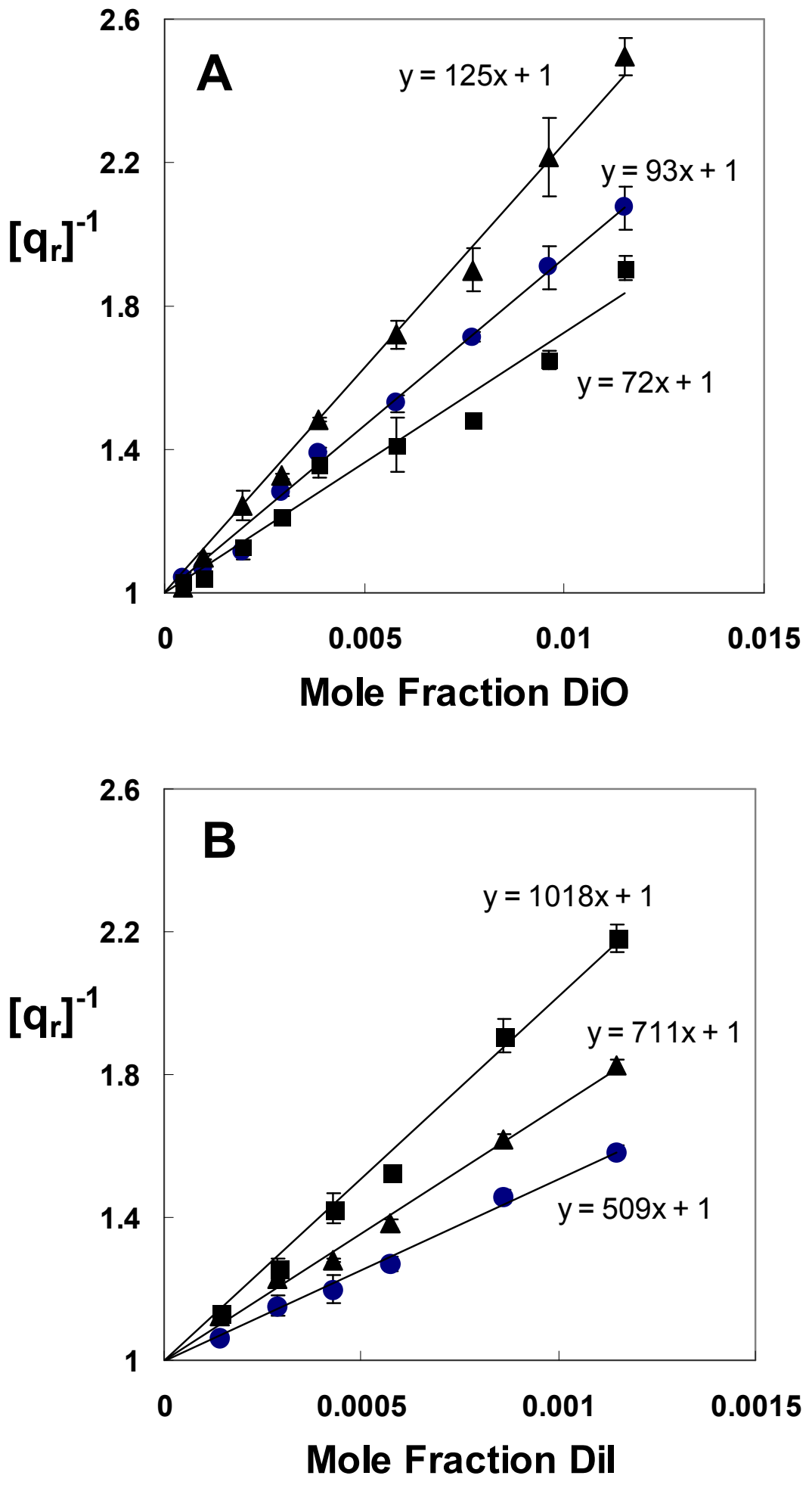

Fig 2. **Acceptor-quenched donor fluorescence, $E(\chi_A)$, also exhibits Stern-Volmer behavior.** Linear S-V quenching curves are easily obtained for both the DHE → DiO (panel A) and DiO → DiI (panel B) probe pairs in all three membrane environments: $L_\alpha$ fluid (●), $L_o$ fluid (▲), and $L_\beta$ gel (■). The DiO→DiI transfer efficiency obeys Eq. 4 for dilute acceptor concentrations only ($\chi_{DiI}$ < 0.0015), but surprisingly, DHE->DiO transfer efficiency actually exhibits S-V behavior all the way up to more than 1.0 mole% DiO. The slope of each line provides the $C_1$ quenching constant for that particular probe pair in the corresponding membrane environment.

Classic quenching plots of the sort shown in Fig. 2 are the easiest means of determining accurately the Stern-Volmer quenching constant for each probe pair and membrane environment. However, these $C_1$ values also conveniently define the maximum-acceptor limit of the LSV regime ($F_A^{Dex} \approx C_0\chi_D\chi_A$, Eq. 6), and since our own studies of membrane phase behavior have been greatly simplified by working within this regime, we have listed in Table 1 the implied LSV-regime limit (taken to be $\chi_A \leq 1/(10\cdot C_1)$) for each probe pair in all three environments.

## COMPARING THE STERN-VOLMER AND WOLBER-HUDSON MODELS

It is informative to compare Eq. 4 with the well-known Wolber-Hudson (W-H) series approximation that describes 2D Forster-quenching of donor fluorescence by randomly distributed acceptor molecules: [7]

$$q_r = \sum_{n=0}^{\infty}\left[\frac{\Gamma\left(n/3+1\right)}{n!}(-\varepsilon C)^n\right] \qquad [7]$$

In this expression, $C \equiv c\cdot R_0^2 = \sigma\cdot\chi_A\cdot R_0^2$ is the "natural" acceptor concentration, where c is the number of acceptor molecules per unit area, $R_0$ is the Forster distance, $\sigma$ represents membrane molecular area, and $\varepsilon \equiv \pi\cdot\Gamma\left(2/3\right)$. The series quickly converges, so that for $C \leq 0.60$ the truncation error is less than 1% by n=12.

At smaller values of $C$, Eq. 7 approaches a form equivalent to Eq. 4,

$$q_r \approx \frac{1}{1+\varepsilon C} \qquad [8]$$

and to illustrate this fact Eqs. 7 and 8 are plotted together in Fig. 3. The two curves agree within ~1.5% for $C \leq 0.10$, so hereafter we will refer to this range of acceptor concentrations as the Wolber-Hudson-Stern-Volmer (WHSV) regime. In terms of acceptor mole fraction, the WHSV regime corresponds to something like $\chi_{Acceptor} \leq 0.001$, if we to assume an efficient probes pair ($R_0$~50 Å) mixing within a typical bilayer-membrane phase (~0.04 Å$^2$/molec). But comparing Eq. 4 with Eq. 8 suggests that $\varepsilon C \approx C_1\chi_A$, a novel correspondence which allows us to compute specific WHSV-regime limits ($\chi_A \leq \varepsilon/(10\cdot C_1)$) for all six combinations of probe pairs and membrane environments (Table 1).

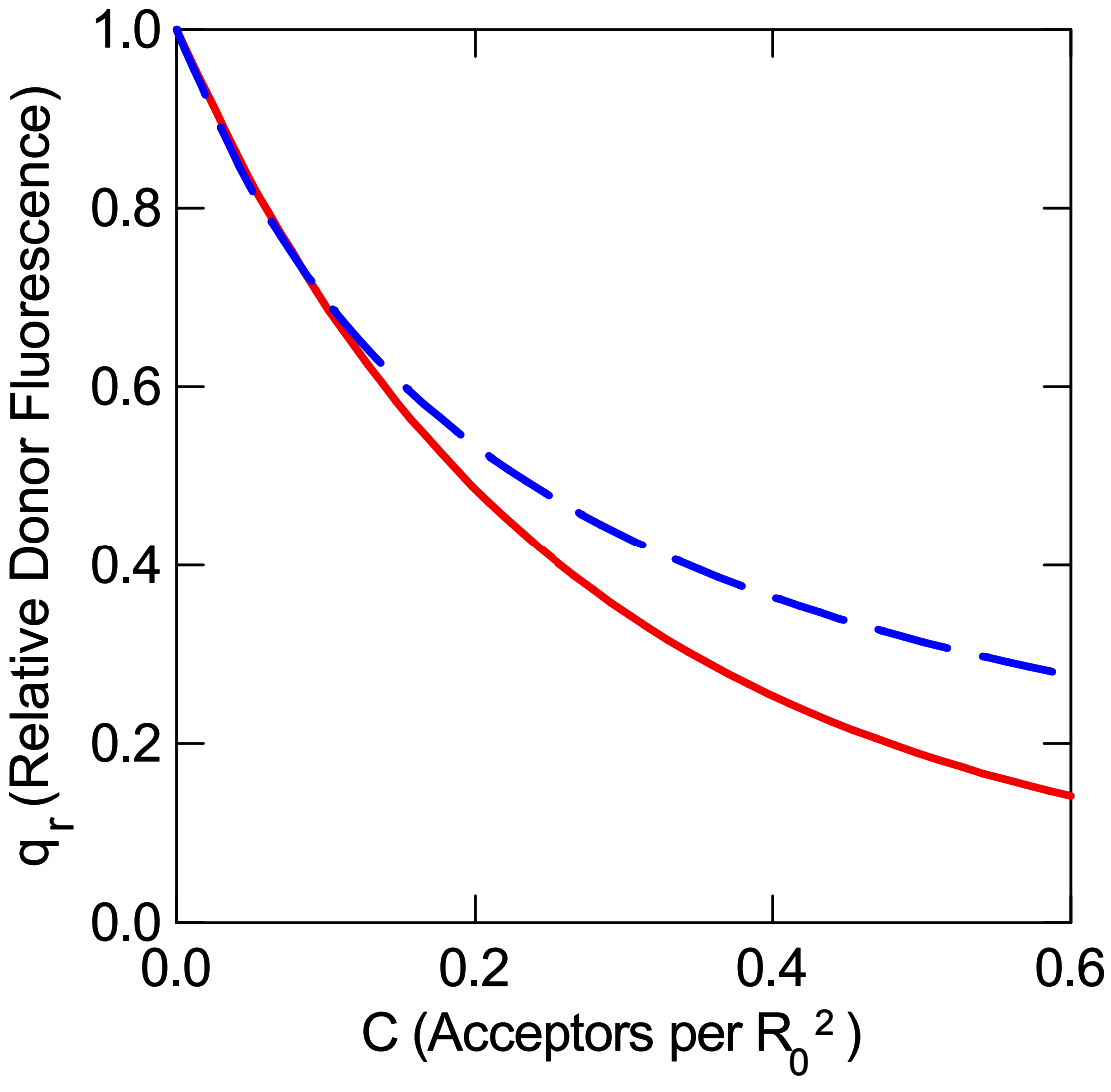

FIG 3. **Convergence of the Wolber-Hudson and Stern-Volmer models in the dilute-acceptor limit.** The close correspondence between Eqs. 7 and 8 for $C \leq 0.10$ (the so-called WHSV regime) suggests that $\varepsilon C \approx C_1\chi_A$.

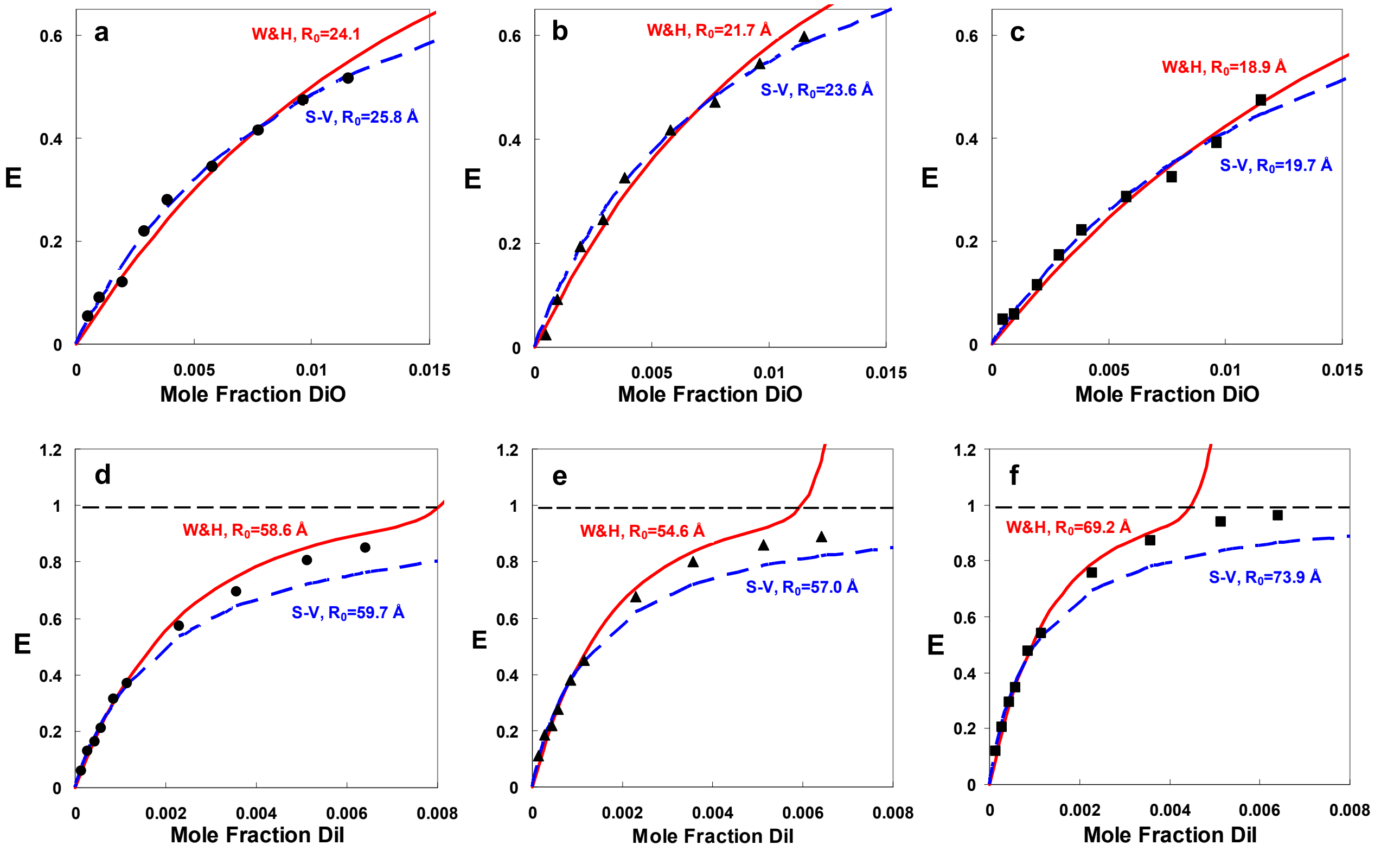

FIG 4. **Full-range comparisons of the relative utility of Stern-Volmer (dashed curve) and Wolber-Hudson (solid curve) expressions for $E(\chi_A)$: Best fits to experimentally determined titration curves for both probe pairs in all three membrane environments.** Experimentally determined acceptor-titration curves are shown for both combinations of probe pairs (DHE→DiO, panels a-c; DiO→DiI, panels d-f) in all three membrane environments ($L_\alpha$ fluid (●), $L_o$ fluid (▲), and $L_\beta$ gel (■)), and the corresponding best-fit $R_0$ values are shown alongside each curve. Both models fit the DHE→DiO data reasonably well over the full range of acceptor concentrations examined, but the W-H fits are slightly skewed, whereas the S-V fits are without any structure in the residuals. Neither model can fit well the DiO→DiI data over the full range of acceptor concentrations (see plots provided in Supplementary Information), so the curves shown in panels d-f represent best fits to dilute-acceptor data only (first 6 data points: $\chi_{DiI} < 0.0015$). When parameterized in this way, both the S-V and W-H expressions eventually fail in the high-DiI regime.

This same correspondence suggests the following interpretation of the Stern-Volmer quenching constant,

$$R_0 \approx \sqrt{\frac{C_1}{\varepsilon\sigma}} \qquad [10].$$

relating each measured $C_1$ value to an effective Forster distance. Of course, $\varepsilon$ is a constant, so by taking appropriate values for $\sigma_{bilayer}$,[14,15,16] we can obtain WHSV estimates of $R_o$ for both our probe pairs in all three membrane environments (Table 1). To the best of our knowledge, Eq. 10 is a novel interpretation of the Stern-Volmer quenching constant in 2D, yet it is clear that all the $R_o$ estimates in Table 1 are in accord with expectations based on spectral-overlap integration.

Even though our primary interest is restricted to FRET in dilute-probe regimes, we also carried out wider-range $E(\chi_A)$ titrations in order to compare more generally the performance of the Stern-Volmer and Wolber-Hudson models. Given our WHSV-regime interpretation of $C_1$ (Eq. 10), both models can be parameterized in terms of $R_o$, so Fig. 4 shows side-by-side comparisons of S-V (dashed curves) and W-H (solid curves) fits to experimentally determined titration data for each probe pair in all three membrane environments. Both models fit the DHE→DiO data reasonably well (Fig 4 a-c), but the W-H fits show some evidence of being skewed, while the S-V fits are without any structure in the residuals. Neither model can be fit to DiO→DiI titration data over the full range of acceptor concentrations (see plots provided in Supplementary Information), so the curves shown in panels d-f represent best fits to dilute-acceptor data only (first 6 data points; $\chi_{DiI} < 0.0015$). When parameterized with dilute-DiI data, both the S-V and W-H models fail in the high-probe regime, though in opposite directions. At higher $\chi_{DiI}$, the S-V model underestimates E, while the W-H model overestimates it.[f]

TABLE I
FRET PARAMETERS DETERMINED FROM S-V QUENCHING CONSTANTS

| Probe Pair | Phase State | $C_1$ † | Maximum $\chi_A$ Limits: WHSV Regime | Maximum $\chi_A$ Limits: LSV Regime | Estimate of $R_o$ (Å) †† |
|---|---|---|---|---|---|
| DHE → DiO | $L_\alpha$ | 93 | $4.7 \times 10^{-3}$ | $1.1 \times 10^{-3}$ | 26 |
| | $L_o$ | 125 | $3.4 \times 10^{-3}$ | $0.80 \times 10^{-3}$ | 24 |
| | $L_\beta$ | 72 | $6.0 \times 10^{-3}$ | $1.4 \times 10^{-3}$ | 20 |
| DiO → DiI | $L_\alpha$ | 509 | $8.5 \times 10^{-4}$ | $2.0 \times 10^{-4}$ | 60 |
| | $L_o$ | 711 | $6.0 \times 10^{-4}$ | $1.4 \times 10^{-4}$ | 57 |
| | $L_\beta$ | 1018 | $4.2 \times 10^{-4}$ | $0.98 \times 10^{-4}$ | 75 |

† from Fig 2.
†† $\sigma_{L_\alpha} = 0.0330$, $\sigma_{L_\beta} = 0.0424$, $\sigma_{L_o} = 0.0513$ Å²/molec (refs 14-16).

## DISCUSSION

It may seem surprising that Wolber and Hudson's model should ever approach Stern-Volmer behavior, since their analysis led to the expected result for a system exhibiting Forster kinetics: non-exponential decay out of the excited donor state:

$$\langle P(t)\rangle = \exp\left[-k_{Dd}t - \varepsilon C(k_{Dd}t)^{1/3}\right] \quad [11]$$

Indeed, the time dependence that Eq. 11 implies for the bimolecular rate coefficient might seem to place Wolber and Hudson's analysis at odds with a simple Stern-Volmer treatment. However, it is in fact well known that Stern-Volmer behavior must always be approached at low quencher concentrations,[17] as may be seen by observing that Eq. 11 approaches a simple exponential as $C$ goes to zero.

Moreover, the substantial body of experimental evidence that we have presented demonstrates clearly that, for reasonably dilute membrane-bound fluorophores, simple Stern-Volmer expressions (Eqs. 3 and 5) do serve perfectly well to describe the concentration-dependence of both the $F_A^{Dex}$ and $E$ steady-state energy-transfer metrics. For any probe pair in any membrane environment, these two S-V expressions can be easily parameterized in terms of $C_1$ (simple experiments of Fig. 2), and this quantity can then be interpreted in terms of an effective Forster distance (Eq. 10). The $C_1$ value can also be used to define conveniently two important steady-state regimes of acceptor concentration: (1) the WHSV regime ($\chi_A \le \varepsilon / (10 \cdot C_1)$), in which Wolber-Hudson quenching (Forster kinetics) is essentially indistinguishable from simple collisional quenching (Stern-Volmer kinetics); and (2) the LSV regime ($\chi_A \le 1/(10 \cdot C_1)$), in which $F_A^{Dex}$ exhibits linear dependence on both donor and acceptor concentrations.

We have ourselves already exploited LSV regimes for both the DHE-DiO and DiO-DiI probe pairs in our FRET-based studies of membrane phase behavior.[10] And given the fact that Eqs. 3 and 5 are considerably more convenient to work with than the W-H series approximation, we conclude that a Stern-Volmer approach—informed by Eq. 10 and the WHSV and LSV regime definitions—should be considered a suitable framework for interpreting FRET experiments between freely diffusing membrane-bound probes.

## CONCLUSIONS

As set forth in the introduction, our purpose here has been to present a simple framework that can serve to relate variations in steady-state FRET metrics to variations in membrane-bound donor and acceptor concentrations. The evidence we have presented confirms simple Stern-Volmer modeling as one such valid framework for dilute-probe experiments. Moreover, our results suggest a simple and systematic three-step strategy for the design of membrane FRET experiments, once suitable probe combinations have been chosen.

First, one should measure the Stern-Volmer quenching constant for each probe combination in each of its anticipated target-membrane environments. This is most easily accomplished by $E(\chi_A)$ titration. Second, one should use the $C_1$ values so obtained to estimate both the effective $R_0$ and the regime-limiting $\chi_A$ values (both WHSV and LSV) for each probe pair in each environment. Third, one should consider all the practical advantages and disadvantages[5] associated with the alternative FRET metrics ($E$ cf. $F_A^{Dex}$), and choose whichever metric most closely suits one's own particular experimental arrangement.

Planning of the larger FRET experiment should then be carried out with all these parameters clearly in mind, balancing signal-intensity considerations (which favor less dilute probes—the WHSV regime) against the advantage offered by a simplest interpretive scheme (favors more dilute probes—the LSV regime). In either case, the fact that both FRET metrics can be expected to exhibit simple behavior within well-defined regimes should aid the development of more effective experimental strategies.

## ACKNOWLEDGMENTS

This work was supported by Research Corporation Award CC6814.

[f] The fact that W-H curves blow up at higher $\chi_A$ is a consequence of the fact that Eq. 7 turns downward (non-physical) for $C > 0.7$.